\documentclass[conference]{IEEEtran}
\IEEEoverridecommandlockouts
\usepackage{cite}
\usepackage{amsmath,amssymb,amsfonts}
\usepackage{algorithmic}
\usepackage{algorithm}
\usepackage{graphicx}
\usepackage{textcomp}
\usepackage{xcolor}
\usepackage{hyperref}
\usepackage{url}
\usepackage[subrefformat=parens]{subcaption}
\usepackage{siunitx}

\usepackage{cleveref}
\crefformat{equation}{Eq.\,(#2#1#3)}
\crefformat{section}{Sec.\,#2#1#3}
\Crefformat{section}{Section~#2#1#3}

\newcommand{\figref}[1]{Fig.\,\ref{#1}}

\newcommand{\secref}[1]{Sec.\,\ref{#1}}

\begin{document}

\title{
Learning Stability of Replay-Based Co-Optimization for Transmission Expansion under Strategic Bidding
}

\author{
\IEEEauthorblockN{Tomonari Kanazawa, Hikaru Hoshino, Eiko Furutani}
\IEEEauthorblockA{\textit{Department of Engineering}, \\
\textit{University of Hyogo} \\
Himeji, Japan \\
\{er25q006@guh, hoshino@eng, furutani@eng\}.u-hyogo.ac.jp}
}

\maketitle

\begin{abstract}

This paper investigates the behavior of learning-based co-optimization for transmission expansion under strategic bidding in electricity markets. In this framework, transmission capacities are updated while market participants simultaneously learn their bidding strategies through deep reinforcement learning, resulting in coupled and non-stationary learning dynamics. We show that transient policy degradation of bidding agents can generate inconsistent cost-capacity samples, which bias the transmission-capacity update and 
prevent the co-optimization process from converging to the desired solution.
To mitigate this, replay-based capacity updates with nearest-neighbor filtering are introduced to exclude inconsistent samples from the update data. We then analyze a new oscillatory behavior that appears when the replay memory is enlarged. Numerical results on the IEEE 30-bus system demonstrate that enlarged replay memories improve robustness against transient policy degradation but can introduce a temporal lag, leading to oscillations unless the capacity-update learning rate is appropriately reduced. These results reveal the trade-off between replay-memory size and learning rate and provide practical guidelines for stable co-optimization.
\end{abstract}


\section{Introduction}

The deregulation of electricity markets and the structural unbundling of generation and transmission sectors have significantly increased the importance of transmission expansion planning under strategic market environments \cite{Sauma2007,Garces2009}. In modern electricity markets, generation companies (GenCos) strategically adjust their bidding behaviors to maximize their own profits, and these behaviors strongly influence market prices, congestion patterns, and ultimately the economic value of transmission investments. As a result, transmission planning and market operation are tightly coupled problems, and it is increasingly important to evaluate transmission expansion while explicitly accounting for strategic bidding behaviors.

To capture strategic bidding behaviors, recent studies have applied multi-agent deep reinforcement learning (MADRL) to electricity market simulations~\cite{Ye2020,Liang2020,Du2021}. 
These approaches model strategic bidding as a sequential decision-making problem and enable market participants to learn bidding policies through repeated interactions with the market environment. At the same time, reinforcement learning has also been explored for transmission expansion and infrastructure planning problems~\cite{MingKui2020,Wang2021,Dong2025}. However, these two research directions have largely developed independently. Existing market-simulation approaches generally assume fixed network configurations, while RL-based expansion-planning studies often neglect the strategic adaptation of market participants.

To bridge this gap, a learning-based co-optimization framework~\cite{Kanazawa2026:PESGM} has recently been proposed in which transmission capacities and bidding strategies are updated simultaneously within a unified learning process. In such frameworks, transmission expansion decisions influence market outcomes and bidding incentives, while the learned bidding behaviors in turn affect the estimated economic value of transmission investments. Although this approach enables more consistent planning under strategic market interactions, it also introduces coupled and non-stationary learning dynamics between operational policies and infrastructure design variables.

One major challenge in this coupled learning process is the instability caused by transient policy degradation during deep reinforcement learning. Because bidding agents occasionally exhibit locally degraded policies during exploration and training, the resulting market outcomes can produce misleading relationships between transmission capacity and electricity cost. In particular, low-cost samples generated by temporarily suboptimal bidding behaviors may incorrectly suggest that insufficient transmission capacity is economically preferable. When such inconsistent samples are directly used for transmission-capacity updates, the co-optimization process can become unstable and fail to converge appropriately.

To improve robustness against these transient failures, a replay-based transmission-capacity update mechanism is introduced in this study. By utilizing accumulated historical samples together with spatial filtering based on nearest-neighbor search, inconsistent samples can be excluded from the update process, thereby stabilizing the capacity learning process against transient policy degradation. However, through long-term simulations, a new oscillatory phenomenon is observed when the replay memory becomes large. While enlarging the replay memory improves robustness by preserving broader historical information, it also introduces temporal lag into the update process due to the persistence of outdated samples. As a result, the transmission-capacity update may exhibit overshoot and oscillatory behavior unless the learning rate is appropriately reduced.

This paper investigates the learning dynamics of replay-based co-optimization for transmission expansion under strategic bidding. In particular, the paper analyzes the relationship between replay-memory size and learning rate, and clarifies the trade-off between robustness against transient policy degradation and oscillatory behavior caused by temporal lag. Numerical experiments on the IEEE 30-bus system demonstrate the proposed update mechanism and provide qualitative guidelines for stable co-optimization under coupled market and infrastructure learning dynamics.

The remainder of this paper is organized as follows. 
\cref{sec:method} introduces the market model and the learning-based co-optimization framework considered in this study. 
\cref{sec:filtering} discusses the instability caused by transient policy degradation and presents the replay-based stabilization method using nearest-neighbor filtering. 
\cref{sec:oscilation} analyzes the oscillatory dynamics induced by replay memory and investigates the trade-off between replay-memory size and learning rate. Finally, \cref{sec:conclusion} concludes the paper.

\section{Learning-based Co-Optimization}
\label{sec:method}

\begin{figure}[t!]
    \centering
    \includegraphics[width=0.99\linewidth]{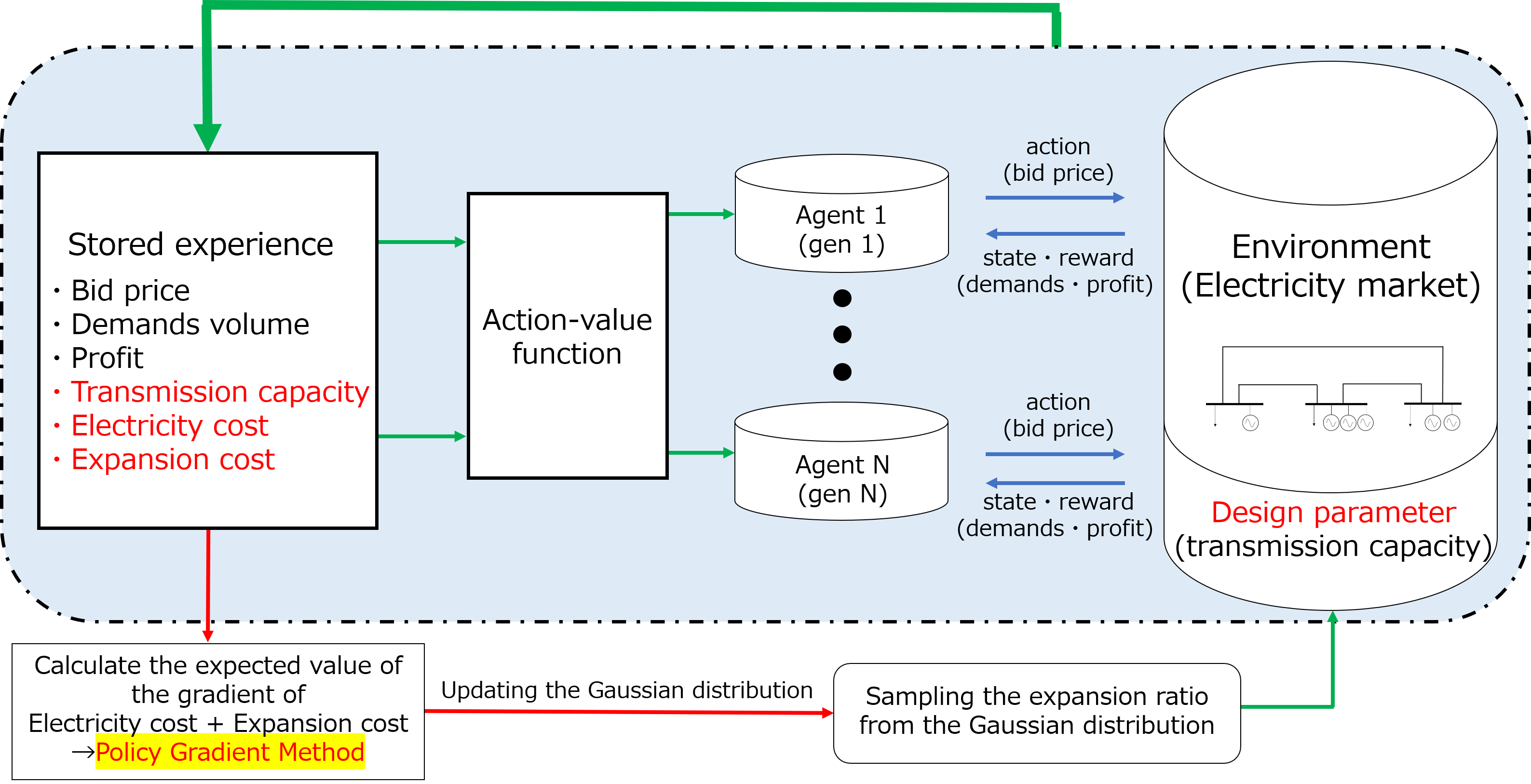}
    \caption{Overview of the co-optimization framework\cite{Kanazawa2026:PESGM}} \label{fig:framework}
\end{figure}

This section briefly reviews the learning-based co-optimization framework proposed in \cite{Kanazawa2026:PESGM}. 
The framework simultaneously learns strategic bidding policies of GenCos and transmission expansion decisions within a unified learning process. 
An overview of the framework is shown in \figref{fig:framework}. 
The upper part represents the multi-agent RL environment used for market simulation, and is described in~\cref{sec:rl_bidding}. 
The lower part illustrates the transmission-capacity update process  as explained in \cref{sec:co-optimization}.

\subsection{Market Simulation Using Multi-agent RL}
\label{sec:rl_bidding}

We consider a day-ahead electricity market where multiple GenCos strategically submit bids to maximize their profits. 
Given the submitted bids, the market operator clears the market using DC optimal power flow (DC-OPF) under power-balance and transmission-capacity constraints. 
The competitive electricity market is modeled as a partially observable multi-agent environment largely based on~\cite{Du2021}. 
Each GenCos is modeled as an autonomous agent that determines its bid price through reinforcement learning. 
At each time step $t$, agent $i$ observes a local state given by
\begin{align}
o_i(t)
=
[
\,P_{\mathrm{load}}(t),\,
\lambda_i^{\mathrm{bid}}(t-1),\,
\omega^\top
\,]^\top,
\end{align}
where $P_{\mathrm{load}}(t)$ stands for the total system demand, $\lambda_i^{\mathrm{bid}}(t-1)$ for the bid submitted at the previous step, and $\omega$ for the transmission-capacity parameters, which is fixed during each episode.
Based on this observation, the agent selects a normalized action $a_i(t)\in(0,1)$, and the bid price is determined as
\begin{align}
\lambda_i^{\mathrm{bid}}(t)
=
\lambda_i^{\mathrm{cost}}
\left(
\alpha_i a_i(t)+1
\right),
\label{eq:bidding_price}
\end{align}
where $\lambda_i^{\mathrm{cost}}$ is the marginal generation cost and $\alpha_i>0$ determines the allowable bid range. 
Larger actions correspond to more aggressive strategic bids. 
After market clearing, each agent receives a reward corresponding to its market profit:
\begin{align}
r_i(t)
=
\left(
\lambda_i^{\mathrm{cleared}}(t)
-
\lambda_i^{\mathrm{cost}}
\right)
P_i^{\mathrm{cleared}}(t),
\end{align}
where $\lambda_i^{\mathrm{cleared}}(t)$ and $P_i^{\mathrm{cleared}}(t)$ represent the cleared market price and dispatched generation output, respectively. 
The agents update their policies to maximize the discounted long-term return
\begin{align}
G_i
=
\sum_t \gamma^t r_i(t),
\end{align}
where $\gamma\in(0,1)$ is the discount factor.
To address the coupled nature of bidding strategies of multiple GenCos, this study employs the Multi-Agent Deep Deterministic Policy Gradient (MADDPG) algorithm, where each agent possesses a decentralized actor network for bidding decisions and a centralized critic network that evaluates joint actions of all agents during training. 
Experience tuples obtained through repeated market interactions are accumulated in replay buffers and used for off-policy updates of actor and critic networks. 

\subsection{Co-optimization of Transmission Capacity}
\label{sec:co-optimization}

To enable simultaneous optimization of transmission capacity and bidding strategies, the transmission-capacity vector
\begin{align}
\omega
=
[\omega_1,\dots,\omega_{N_\ell}]^\top
\end{align}
is treated as a learnable design variable updated together with the market policies, where let  $N_\ell$ is the number of transmission lines considered for capacity expansion. 
At the beginning of each episode in the RL training loop, the transmission capacities are sampled from a stochastic design policy parameterized by $\mu$:
\begin{align}
p_{\mu}(\omega)
=
\prod_{\ell = 1}^{N_\ell}
p_{\mu_\ell}(\omega_\ell),
\end{align}
where $\mu_\ell$ denotes the learnable parameter associated with line $\ell$.
Each design variable corresponds to the capacity increment $\Delta L_\ell$ of line $\ell$. 
The design policy is modeled using a Gaussian distribution:
\begin{align}
p_{\mu_\ell}(\omega_\ell)
=
\frac{1}{\sqrt{2\pi\sigma_\ell^2}}
\exp
\left(
-
\frac{
(\omega_\ell-\mu_\ell)^2
}{
2\sigma_\ell^2
}
\right),
\end{align}
where $\sigma_\ell$ is a fixed standard deviation controlling exploration during learning.

The objective of the co-optimization is to minimize the total system cost consisting of operational cost and transmission investment cost. 
Accordingly, the episodic return for transmission planning is defined as
\begin{align}
G_{\mathrm{total}}
=
-
\left(
W_{\mathrm{anu}}
\sum_{t=1}^{T}
C_{\mathrm{oper}}(t)
+
C_{\mathrm{exp}}
\right),
\end{align}
where $C_{\mathrm{oper}}(t)$ is the operational market cost at time step $t$, ${T}$ is the total number of time steps,  $C_{\mathrm{exp}}$ is the annualized transmission expansion cost, and $W_{\mathrm{anu}}$ is the annualization factor.
Based on the sampled market outcomes, the transmission-design parameter $\mu$ is updated using a policy-gradient method:
\begin{align}
\nabla_{\mu}
\mathbb{E}[G_{\mathrm{total}}]
=
\mathbb{E}
\left[
\nabla_{\mu}
\ln p_{\mu}(\omega)
\left(
G_{\mathrm{total}}
-
\bar{G}_{\mathrm{total}}
\right)
\right],
\label{eq:update}
\end{align}
where $\bar{G}_{\mathrm{total}}$ is a moving-average baseline introduced to reduce the variance of gradient estimation.
In~\cite{Kanazawa2026:PESGM}, the expectation in \cref{eq:update} was approximated using only the most recent episodes. 
In contrast, this study investigates a replay-memory-based update method, where accumulated historical samples are reused for estimating the gradient of the transmission-capacity update.

\section{Replay-Based Stabilization of Co-Optimization}
\label{sec:filtering}

This section presents the replay-based stabilization method for the co-optimization framework introduced above. 
The IEEE 30-bus market model introduced in \cite{Kanazawa2026:PESGM} is used to demonstrate how replay-based capacity updates improve robustness against transient policy degradation of bidding agents.

\subsection{Simulation Settings}

\begin{figure}
    \centering
    \includegraphics[width=0.85\linewidth]{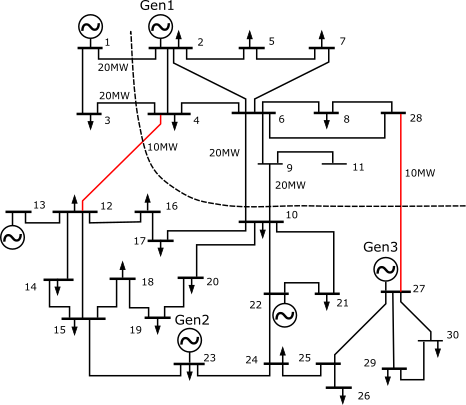}
    \caption{IEEE 30-bus system diagram}
    \label{fig:IEEE_30bus_system}
\end{figure}

The configuration of the IEEE 30-bus system is shown in \figref{fig:IEEE_30bus_system}.
Generators connected to buses 2 (Gen1), 23 (Gen2), and 27 (Gen3) are modeled as strategic bidding agents, while all remaining generators are assumed to bid at their marginal costs. 
The marginal cost of these agents is \$50/MWh, and they determine their bid prices (up to twice the marginal cost) to maximize their own profits. The other generators have a marginal cost of \$55/MWh and are assumed to bid truthfully at that price.
In this system, a relatively large demand is concentrated in the upper-right area, and the targets for transmission capacity optimization are the lines between buses 4-12 and 27-28, both with an initial capacity of 10 MW, and an expansion cost of $\$100,000\mathrm{/MW/year}$.
Market simulations are performed with an hourly resolution over $T=48$ time steps.
As reported in \cite{Kanazawa2026:PESGM}, Gen1 possesses significant market power under the considered market conditions, while increasing the capacity of line 4--12 is particularly effective in reducing the total system cost.
Although transmission expansion mitigates congestion, preliminary simulations showed that the profit-maximizing bid of Gen1 remains around \$90/MWh, indicating that strategic bidding incentives persist even after capacity expansion.
Accordingly, the following analysis focuses on how the learning dynamics of Gen1's bidding policy affect the capacity update of line 4--12 during the co-optimization process.


\subsection{Instability in the Conventional Approach} \label{sec:instability}

\begin{figure}[t]
\centering
\begin{subfigure}{\linewidth}
    \centering
    \includegraphics[width=0.9\linewidth]
    {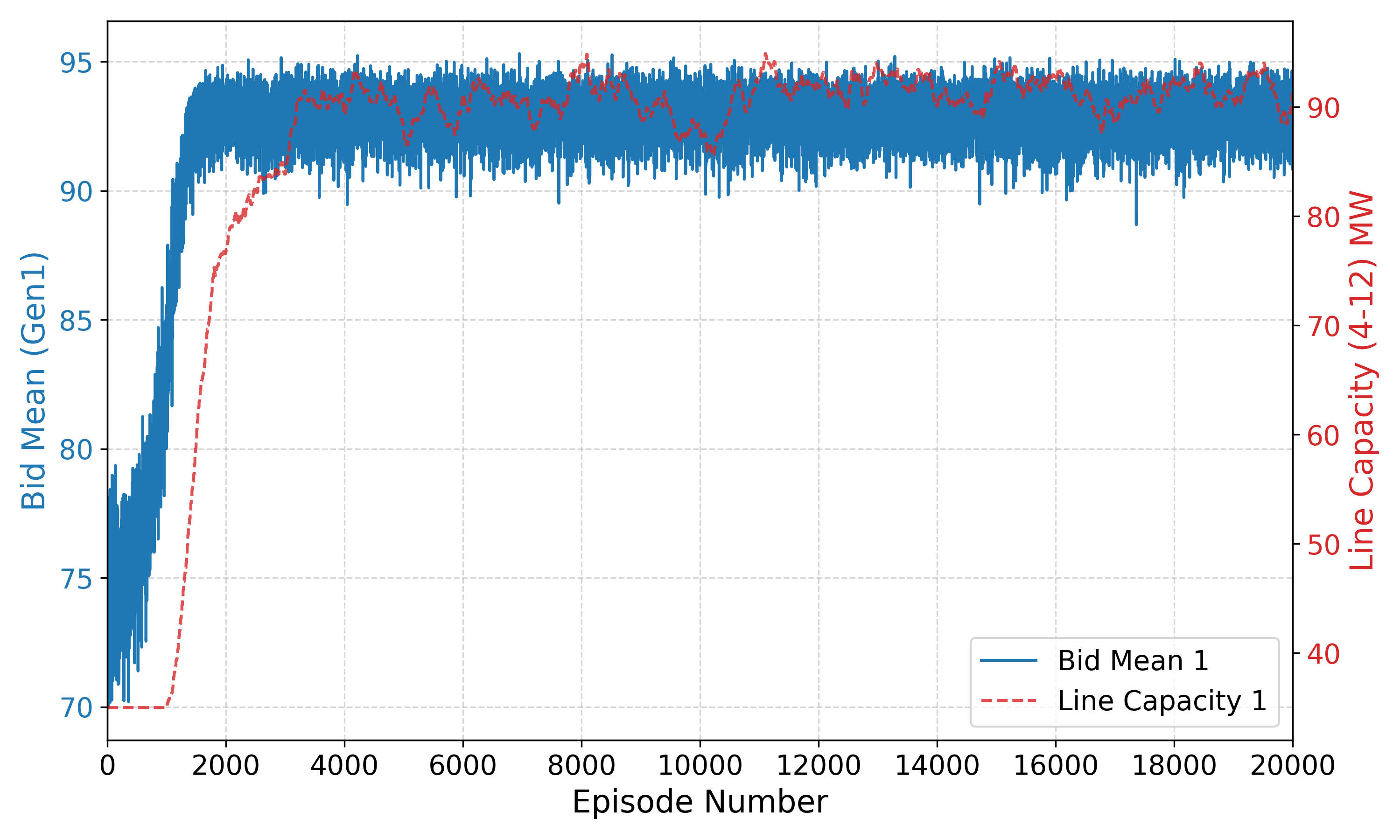}
    \caption{Successful case}
    \label{fig:bid_cap_success_latest50}
\end{subfigure}
\vspace{2mm}
\begin{subfigure}{\linewidth}
    \centering
    \includegraphics[width=0.9\linewidth]
    {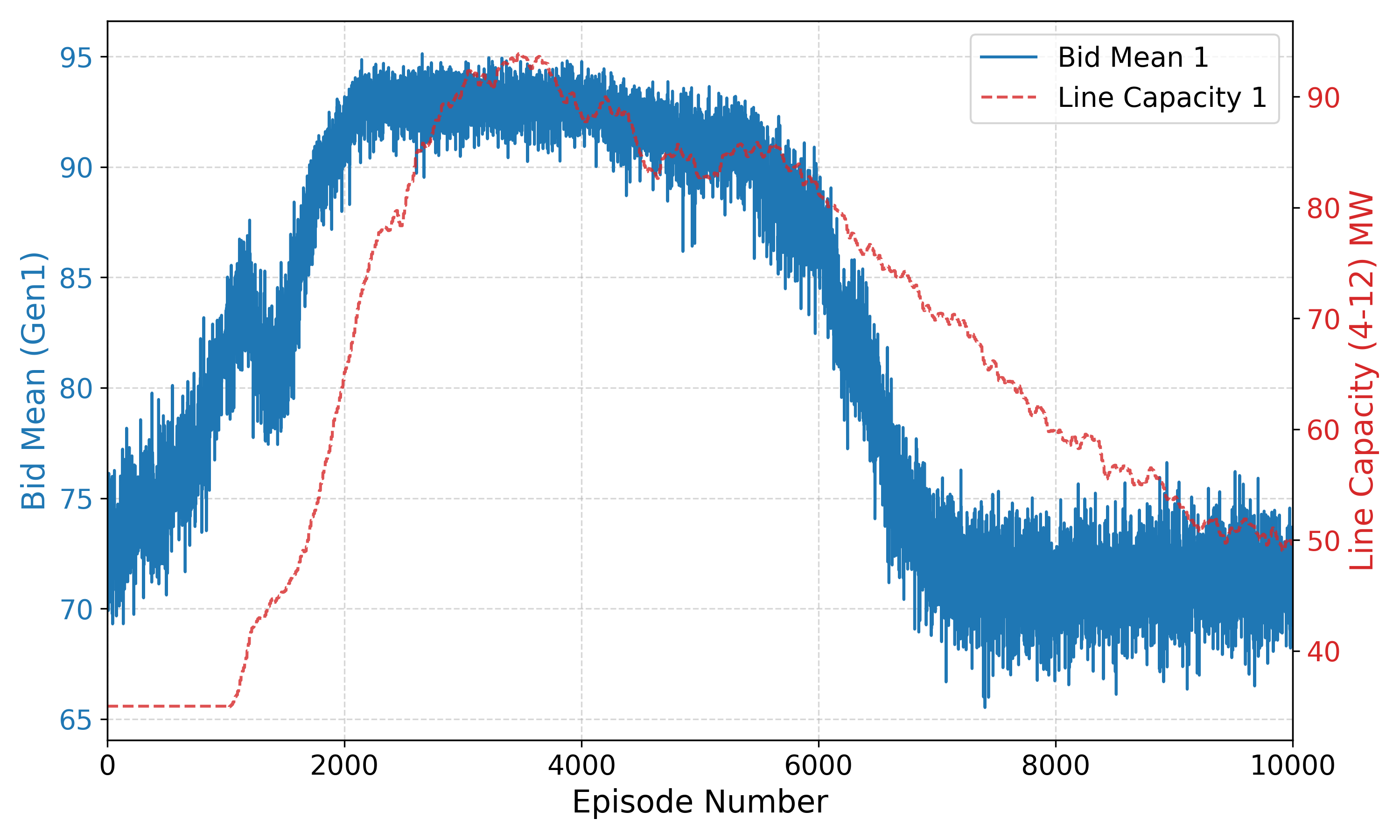}
    \caption{Failure caused by transient policy degradation}
    \label{fig:bid_cap_fail_latest50}
\end{subfigure}
\caption{Learning trajectories with the conventional capacity update method}
\label{fig:bid_cap_latest50}
\end{figure}

During the co-optimization process, transient policy degradation of bidding agents can directly affect the transmission-capacity update and 
prevent the co-optimization process from converging to the desired solution.
To illustrate this phenomenon, \figref{fig:bid_cap_latest50} shows representative learning trajectories obtained using the conventional update method, which estimates the capacity-update gradient from only the most recent 50 episodes. 
The left axis indicates the average bid price of Gen1, while the right axis shows the transmission capacity of line 4--12. In the successful case (\figref{fig:bid_cap_success_latest50}), Gen1 learns a near-optimal bidding strategy around \$90/MWh and maintains it throughout training, resulting in a stable transmission-capacity update. 
In contrast, in the failure case (\figref{fig:bid_cap_fail_latest50}), the bidding policy initially converges to a similar level but subsequently deteriorates after approximately 5,000 episodes. As the bid price decreases, the transmission capacity is also driven downward, eventually leading to an underestimated expansion level. This result suggests that temporary learning failures of bidding agents can propagate to the transmission-planning layer and destabilize the co-optimization process.

\begin{figure}[t]
\centering
\begin{subfigure}{\linewidth}
    \centering
    \includegraphics[width=0.77\linewidth]
    {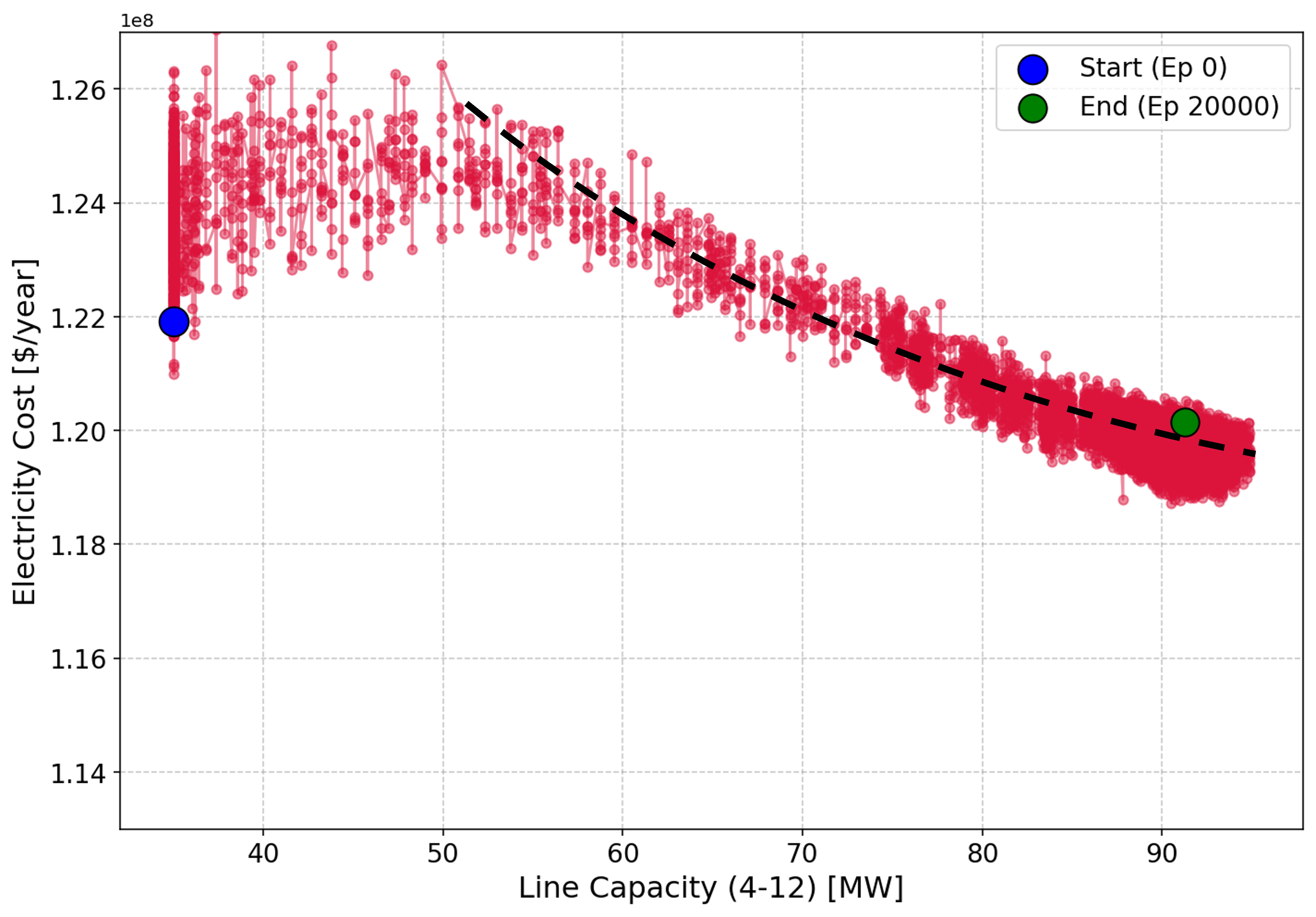}
    \caption{Successful case}
    \label{fig:cap_cost_success_latest50}
\end{subfigure}
\vspace{2mm}
\begin{subfigure}{\linewidth}
    \centering
    \includegraphics[width=0.77\linewidth]
    {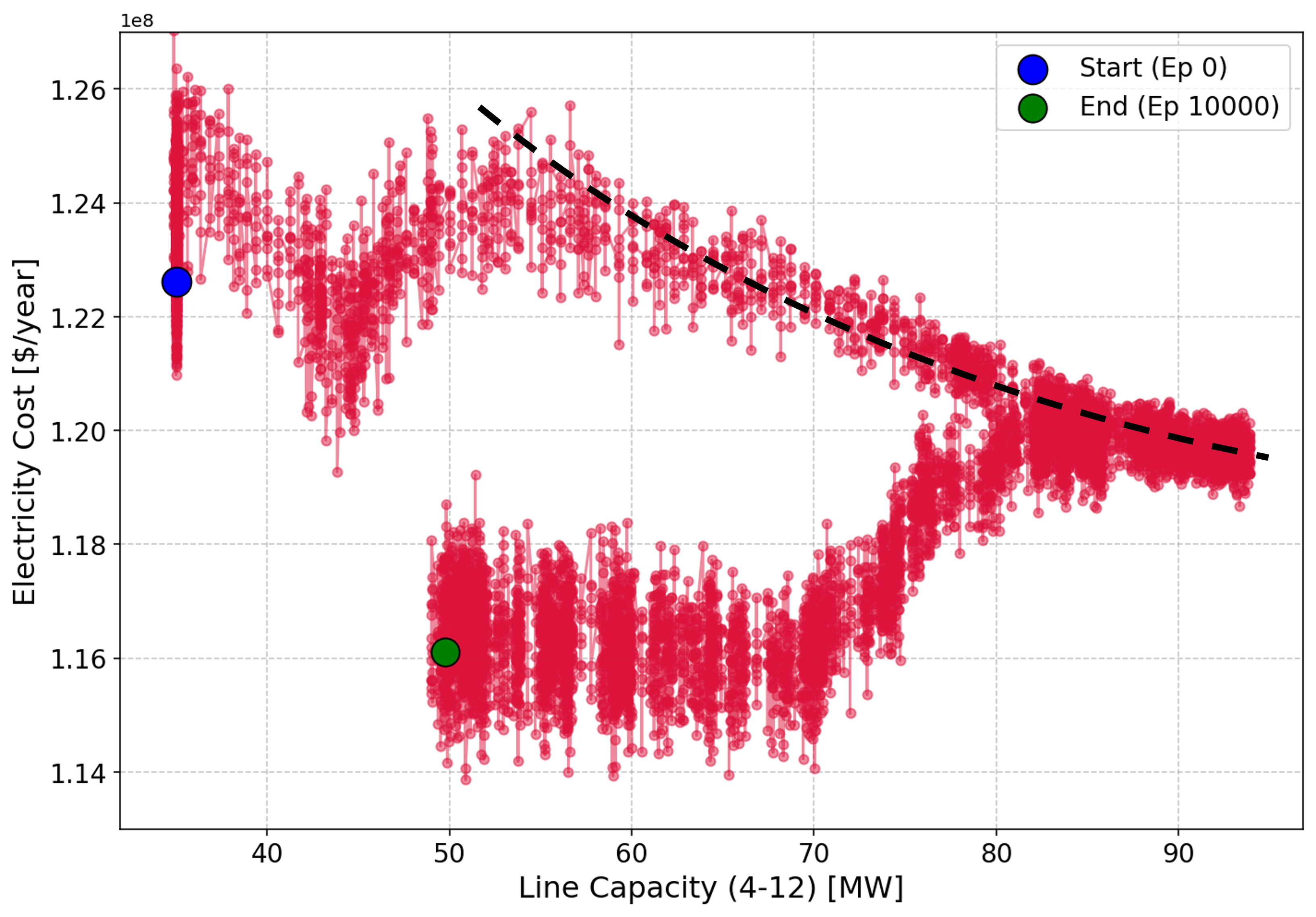}
    \caption{Failure caused by transient policy degradation}
    \label{fig:cap_cost_fail_latest50}
\end{subfigure}
\caption{Relationship between the transmission capacity and the resulting total electricity cost}
\label{fig:cap_cost_latest50}
\end{figure}

To understand the mechanism behind this failure, 
\figref{fig:cap_cost_latest50} shows the relationship between the transmission capacity of line 4--12 and the resulting total electricity cost for the successful and failure cases, respectively. In the successful case (\figref{fig:cap_cost_success_latest50}), the samples are distributed around the monotonically decreasing cost-capacity relationship indicated by the dashed line, where increasing transmission capacity alleviates congestion and reduces the total system cost. 
Consequently, the capacity-update process correctly captures the economic value of transmission expansion. 
In the failure case (\figref{fig:cap_cost_fail_latest50}), however, a large number of samples appear significantly below this expected trend. 
These samples are generated when Gen1 temporarily adopts suboptimal low-price bids, yielding 
spuriously
low electricity costs even under insufficient transmission capacity. 
Because the conventional method relies only on recent samples, such inconsistent data can strongly bias the estimated capacity gradient and lead to inappropriate capacity reductions. Therefore, improving the robustness of the transmission-capacity update against transient policy degradation is a key challenge for stable co-optimization.

\subsection{Replay-Based Capacity Update with Sample Filtering}

To address the instability discussed in \secref{sec:instability}, we introduce a replay-based update mechanism for transmission-capacity learning. Instead of estimating the capacity-update gradient using only the most recent episodes, past co-optimization experiences are accumulated in a replay memory and reused for gradient estimation. This idea is analogous to experience replay in RL~\cite{Mnih15:DQN,Horgan2018:Ape-X}.
However, simply reusing past samples is insufficient because replay memory may contain data generated during temporary policy degradation of bidding agents. 
To reduce their influence, a sample-filtering procedure is introduced before constructing the update minibatch. 
For each sample in the replay memory, neighboring samples are identified in the transmission-capacity space using a nearest-neighbor search implemented with a KD-tree.
The electricity costs of these neighboring samples are compared, and only the sample with the highest electricity cost within the neighborhood is retained. This procedure preferentially removes samples associated with transient policy degradation while preserving samples that better represent the underlying cost-capacity relationship. The minibatch used for the transmission-capacity update is then randomly sampled from the remaining filtered data.



\begin{figure}[t]
\centering
\begin{subfigure}{\linewidth}
    \centering
    \includegraphics[width=0.9\linewidth]
    {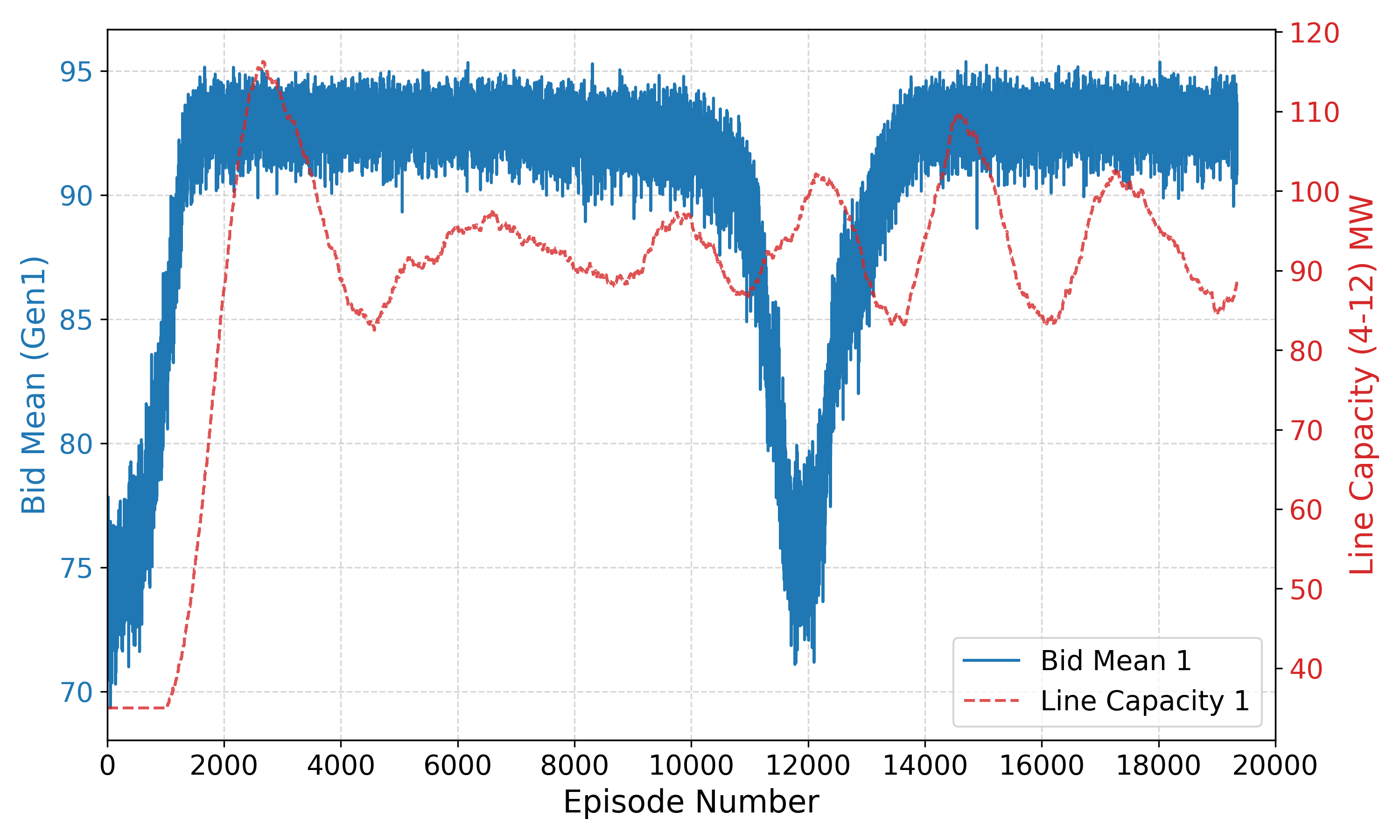}
    \caption{Learning trajectory with replay-based update and sample filtering.}
    \label{fig:bid_cap_KDTree}
\end{subfigure}
\vspace{2mm}
\begin{subfigure}{\linewidth}
    \centering
    \includegraphics[width=0.77\linewidth]
    {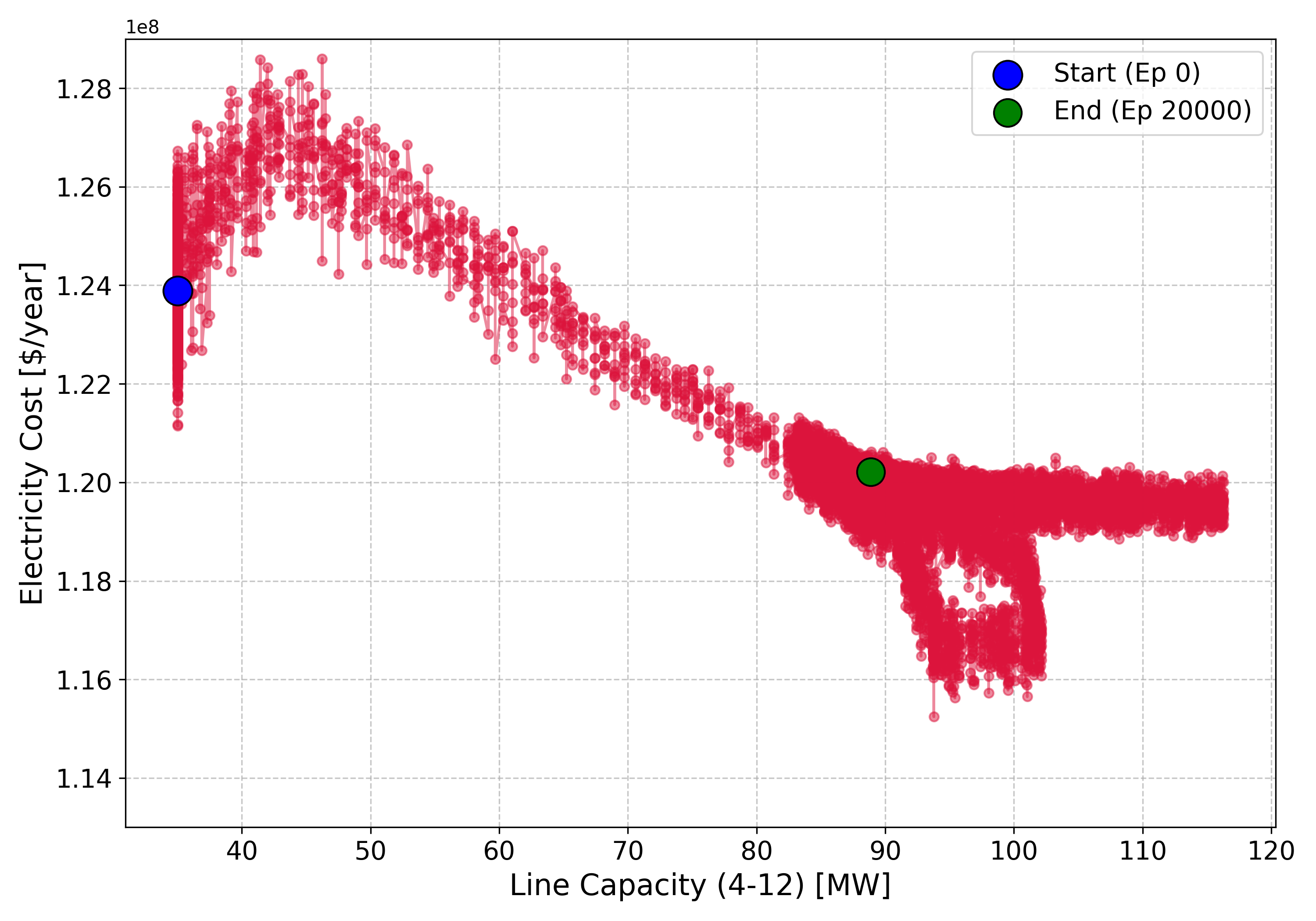}
    \caption{Cost-capacity relationship after sample filtering.}
    \label{fig:elec_cost_KDTree}
\end{subfigure}
\caption{Effect of the proposed replay-based update with sample filtering}
\label{fig:KDTree_results}
\end{figure}

The effectiveness of the proposed method is evaluated using the same simulation setting as in \secref{sec:instability}.
The replay memory size is set to 2000 episodes, the number of nearest neighbors is set to $k=5$, and the transmission-capacity update is performed using mini-batches of 50 samples. The resulting learning trajectory and cost-capacity relationship are shown in \figref{fig:KDTree_results}.
As shown in \figref{fig:bid_cap_KDTree}, a transient policy degradation of Gen1 occurs around episodes 11,000--13,000, where the bid price temporarily drops from its near-optimal level. 
Unlike the conventional method shown in \figref{fig:bid_cap_fail_latest50}, however, the transmission capacity remains within an appropriate range and does not exhibit the severe degradation observed previously. 
After this transient degradation, the agent recovers its near-optimal bidding policy.
Furthermore, \figref{fig:elec_cost_KDTree} shows that the filtered samples largely follow the expected
cost-capacity relationship, in contrast to the substantial deviations observed in
\figref{fig:cap_cost_fail_latest50}.
As a result, the transmission-capacity update becomes less sensitive to temporary policy degradation of bidding agents, preventing inappropriate capacity reductions.

The computational overhead of the replay-based capacity update is small relative to the MADDPG training.
In our implementation, the replay-based capacity update requires 7.2--\SI{30.4}{ms} for replay-memory sizes of $W=2000$--7000.
The main components are KD-tree construction (0.53--\SI{2.33}{ms}), $k$-nearest-neighbor search (2.61--\SI{11.40}{ms}), and sample filtering (2.55--\SI{10.90}{ms}). 
For comparison, the MADDPG learner requires approximately \SI{5.4}{ms} per update, corresponding to approximately \SI{780}{ms} per episode with three agents and 48 time steps.
Since the capacity update is performed once every 10 episodes, its average computational cost is less than 0.4\% of the MADDPG learner-update time even for $W=7000$.

\section{Oscillatory Dynamics and Stability Trade-off}
\label{sec:oscilation}

The replay-based update with sample filtering successfully eliminates the capacity underestimation caused by transient policy degradation of bidding agents. Nevertheless, the learning trajectories in \figref{fig:KDTree_results} reveal that the transmission capacity does not converge monotonically to its equilibrium value. Instead, after approaching the economically appropriate level, the capacity repeatedly overshoots and undershoots around the equilibrium. Even during periods in which the bidding policies exhibit stable learning behavior, the transmission capacity continues to oscillate, and the observed oscillation cannot be attributed to the transient policy degradation discussed in \secref{sec:filtering}.

The source of this oscillation can be understood from the role of replay memory in the transmission-capacity update. By accumulating past experiences, replay memory provides a richer set of samples for estimating the relationship between transmission capacity and electricity cost. This improves robustness against transient policy degradation and enables more reliable capacity updates. However, older samples remain influential for a longer period of time. As the transmission capacity changes during learning, these historical samples no longer accurately represent the current operating condition, introducing a temporal lag into the update process. Consequently, the planner may continue increasing the transmission capacity even after the economically appropriate level has been reached, resulting in overshoot. Subsequent corrective updates then reduce the capacity, producing repeated overshoot-and-correction cycles that appear as oscillatory dynamics.

\begin{figure}[t]
\centering
\begin{subfigure}{\linewidth}
    \centering
    \includegraphics[width=0.85\linewidth]{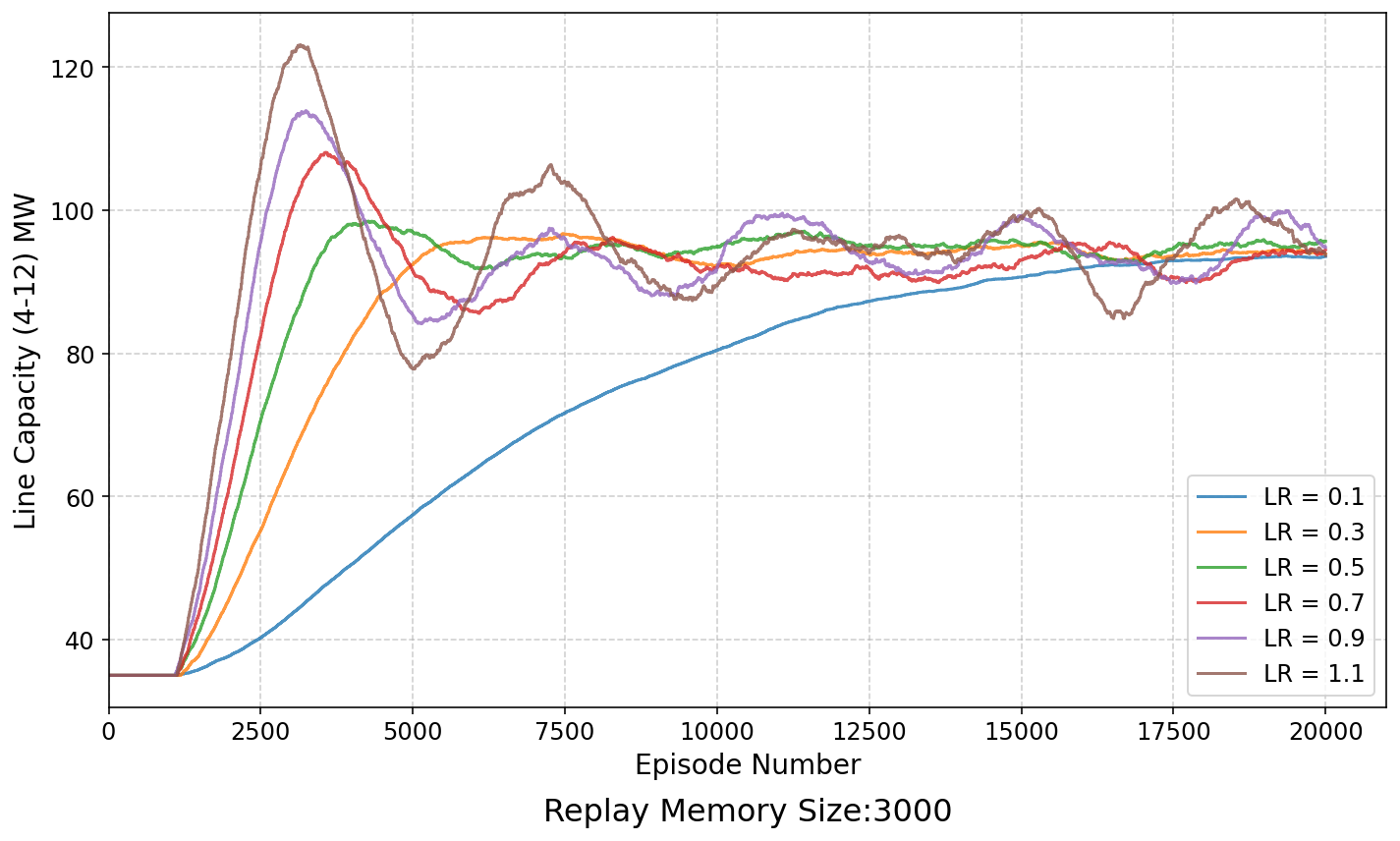}
    \caption{Replay memory size: 3000}
    \label{fig:comp_cap_rl_3000}
\end{subfigure}
\vspace{2mm}
\begin{subfigure}{\linewidth}
    \centering
    \includegraphics[width=0.85\linewidth]{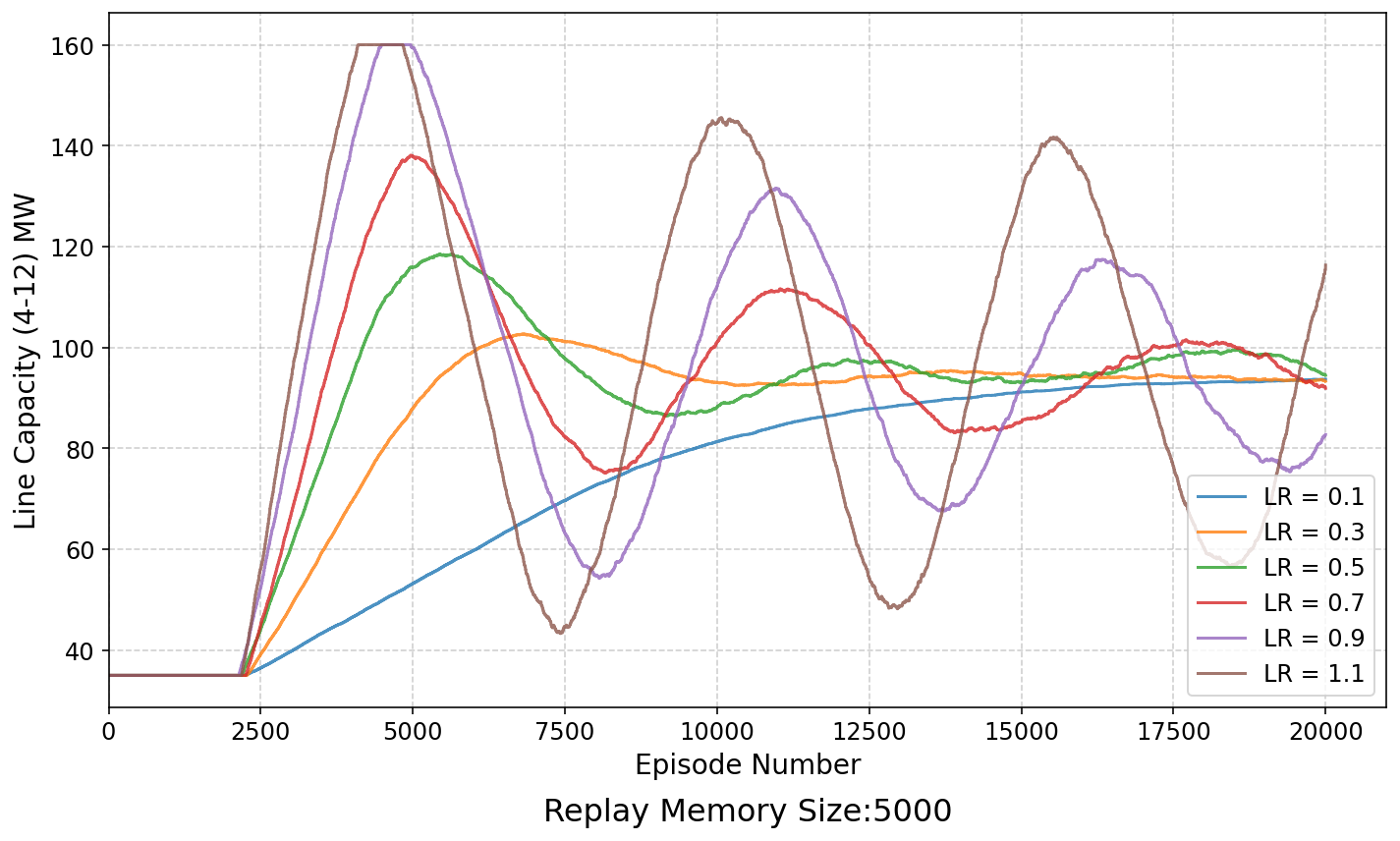}
    \caption{Replay memory size: 5000}
    \label{fig:comp_cap_rl_5000}
\end{subfigure}
\vspace{2mm}
\begin{subfigure}{\linewidth}
    \centering
    \includegraphics[width=0.85\linewidth]{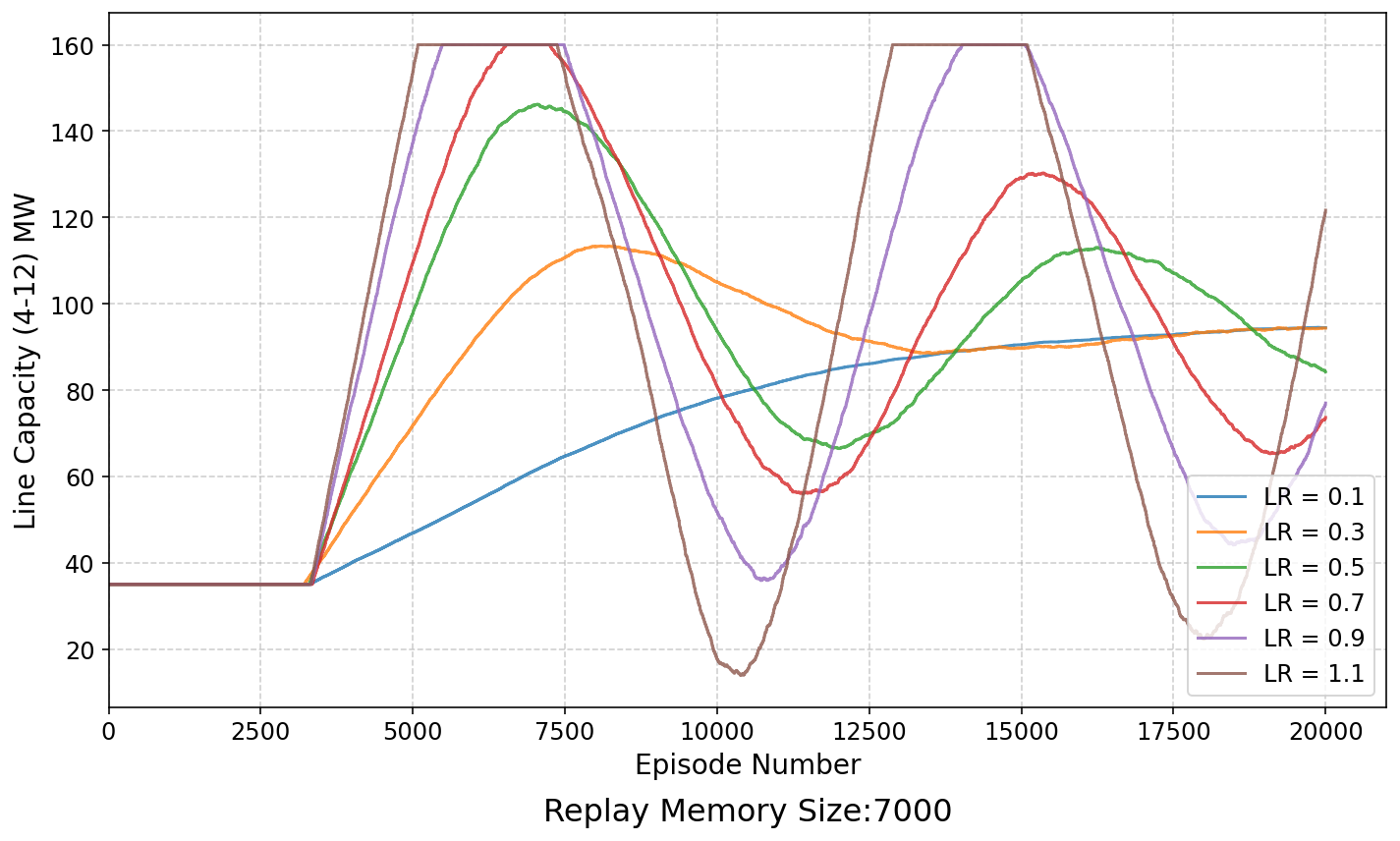}
    \caption{Replay memory size: 7000}
    \label{fig:comp_cap_rl_7000}
\end{subfigure}
\caption{Influence of replay-memory size and learning rate}
\label{fig:comp_cap_rl}
\end{figure}

To investigate this mechanism quantitatively, the replay-memory size and the learning rate of the transmission-capacity update are varied systematically. The resulting learning trajectories are shown in \figref{fig:comp_cap_rl}. 
Comparing the results across different replay-memory sizes reveals that increasing the replay-memory size generally amplifies the oscillation and increases the overshoot magnitude. This tendency is consistent with the interpretation discussed above: a larger replay memory retains a greater number of older experiences, causing the learning process to rely on information that reflects past network conditions. As a result, the transmission-capacity update responds more slowly to the current state of the system, 
thereby introducing a larger temporal lag and making oscillatory behavior more pronounced. In contrast, reducing the learning rate suppresses the oscillation under all replay-memory settings. Because the transmission capacity is updated more conservatively at each iteration, abrupt changes are avoided and the learning dynamics become more stable, leading to smaller oscillations and reduced overshoot.

These results reveal a fundamental trade-off in replay-based co-optimization. Increasing the replay-memory size improves the representation of the cost–capacity relationship by incorporating a broader range of past experiences and enhances robustness against transient policy degradation. 
However, it also introduces a longer temporal lag between the current system state and the information used for capacity updates, which amplifies oscillatory behavior. 
Consequently, larger replay memories require smaller learning rates to maintain stable learning, although this inevitably slows the convergence of the transmission-capacity update. 
Overall, these findings provide a practical guideline for hyperparameter selection and highlight the importance of jointly tuning replay memory and capacity-update dynamics in the proposed framework.

\section{Conclusion}
\label{sec:conclusion}

This paper investigated the learning dynamics of replay-based co-optimization for transmission expansion under strategic bidding in electricity markets. Numerical studies on the IEEE 30-bus system showed that transient policy degradation of bidding agents can generate inconsistent cost-capacity samples and 
prevent the co-optimization process from converging to the desired solution. To address this issue, a replay-based update mechanism with KD-tree-based sample filtering was introduced. The proposed method successfully improved robustness against transient policy degradation, while also revealing a new oscillatory behavior induced by replay memory. Further analysis clarified a trade-off between replay-memory size and learning rate: larger replay memories improve robustness but require more conservative capacity updates to maintain stable learning.

Future work will focus on extending the proposed framework to more practical transmission-expansion settings, including discrete investment decisions and multiple candidate expansion projects. Another important direction is the application of the framework to larger-scale electricity-market environments, enabling co-optimization of transmission planning and strategic bidding under more realistic operating conditions.

\section*{Acknowledgment}

This work was supported in part by Grant-in-Aid for
Scientific Research (KAKENHI) from the Japan Society
for Promotion of Science (\#26K07557) and by Japan Power Academy (2025-H19).

\bibliographystyle{ieeetr}
\bibliography{ref}


\end{document}